\documentclass[aps,prl,amstext,amsmath,twocolumn,superscriptaddress,showpacs]{revtex4-1}

\usepackage{graphicx}
\usepackage{color}

\newcommand{\Fref}[1]{Figure~\ref{#1}}
\newcommand{\fref}[1]{Fig.~\ref{#1}}

\newcommand{\dlt}{$\Delta \lambda (T)$}

\newcommand{\bak}{(Ba$_{1-x}$K$_x$)Fe$_2$As$_2$}

\newcommand{\spm}{$s_{\pm}$}

\newcommand{\tc}{$T_c$}

\begin{document}

\title{Nodeless multiband superconductivity in stoichiometric single crystalline CaKFe$_4$As$_4$}

\author{Kyuil Cho}
\affiliation{Ames Laboratory, Ames and Department of Physics $\&$ Astronomy, Iowa State University, Ames, IA 50011, U.S.A.}

\author{A.~Fente}
\affiliation{Laboratorio de Bajas Temperaturas, Unidad Asociada UAM CSIC, Departamento de F\'isica de la Materia Condensada, Instituto Nicol\'as Cabrera and Condensed Matter Physics Center (IFIMAC), Universidad Aut\'onoma de Madrid, Spain}

\author{S.~Teknowijoyo}
\affiliation{Ames Laboratory, Ames and Department of Physics $\&$ Astronomy, Iowa State University, Ames, IA 50011, U.S.A.}

\author{M.~A.~Tanatar}
\affiliation{Ames Laboratory, Ames and Department of Physics $\&$ Astronomy, Iowa State University, Ames, IA 50011, U.S.A.}

\author{T.~Kong}
\affiliation{Ames Laboratory, Ames and Department of Physics $\&$ Astronomy, Iowa State University, Ames, IA 50011, U.S.A.}

\author{W.~Meier}
\affiliation{Ames Laboratory, Ames and Department of Physics $\&$ Astronomy, Iowa State University, Ames, IA 50011, U.S.A.}

\author{U. Kaluarachchi}
\affiliation{Ames Laboratory, Ames and Department of Physics $\&$ Astronomy, Iowa State University, Ames, IA 50011, U.S.A.}

\author{I.~Guillam\'{o}n}
\affiliation{Laboratorio de Bajas Temperaturas, Unidad Asociada UAM CSIC, Departamento de F\'isica de la Materia Condensada, Instituto Nicol\'as Cabrera and Condensed Matter Physics Center (IFIMAC), Universidad Aut\'onoma de Madrid, Spain}

\author{H.~Suderow}
\affiliation{Laboratorio de Bajas Temperaturas, Unidad Asociada UAM CSIC, Departamento de F\'isica de la Materia Condensada, Instituto Nicol\'as Cabrera and Condensed Matter Physics Center (IFIMAC), Universidad Aut\'onoma de Madrid, Spain}

\author{S.~L.~Bud'ko}
\affiliation{Ames Laboratory, Ames and Department of Physics $\&$ Astronomy, Iowa State University, Ames, IA 50011, U.S.A.}

\author{P.~C.~Canfield}
\affiliation{Ames Laboratory, Ames and Department of Physics $\&$ Astronomy, Iowa State University, Ames, IA 50011, U.S.A.}

\author{R.~Prozorov}
\email[Corresponding author: ]{prozorov@ameslab.gov}
\affiliation{Ames Laboratory, Ames and Department of Physics $\&$ Astronomy, Iowa State University, Ames, IA 50011, U.S.A.}

\date{20 June 2016}

\begin{abstract}
Measurements of the London penetration depth, \dlt, and tunneling conductance in single crystals of the recently discovered stoicheometric,  iron - based superconductor, CaKFe$_4$As$_4$ (CaK1144) show nodeless, two effective gap superconductivity with a larger gap of about 6-9 meV and a smaller gap of about 1-4 meV. Having a critical temperature, $T_{c,onset} \approx$ 35.8 K, this material behaves similar to slightly overdoped \bak\ (e.g. $x=$0.54, $T_c \approx$ 34 K)---a known multigap $s_{\pm}$ superconductor. We conclude that the superconducting behavior of stoichiometric CaK1144 demonstrates that two-gap $s_{\pm}$ superconductivity is an essential property of high temperature superconductivity in iron - based superconductors, independent of the degree of substitutional disorder.
\end{abstract}
\maketitle

Iron-based superconductors (IBS) are represented by a diverse group of different structural families all containing iron layers, which are believed to play the key - role in superconductivity with the superconducting transition temperature, \tc, ranging from 2 to 56 K \cite{Mazin2010,ChubukovAR2012,Hirschfeld2016}. Most of these compounds contain fractional amounts of different ions forming superconducting ``domes" as a function of the composition, resulting in complex phase diagrams and very rich physics \cite{Johnston2010review,Paglione2010review,Canfield2010review122, StewartRMP2011}. The highest T$_c$ is found at fractional compositions, which unavoidably  have finite degrees of substitutional disorder. This represents a serious problem in understanding the pairing mechanism that is ultimately responsible for the high $T_c$'s, because, in materials with anisotropic or sign - changing gaps, any disorder represents extra difficulty to quantify pair-breaking effects, in addition to non-spin-flip scattering \cite{EfremovPRB2011,Kogan2009}. Among the few of stoichiometric IBS, KFe$_2$As$_2$ ($T_c \approx$ 3.6 K), LiFeAs ($T_c \approx$ 18 K), FeSe ($T_c \approx$ 9 K in bulk crystals at ambient pressure) and FeS ($T_c \approx$ 5 K), the recently discovered CaKFe$_4$As$_4$ (CaK1144) clearly stands out with a substantially higher value of $T_{c,onset} \approx$ 35.8 K and $H_{c2,c} \approx$71 T \cite{Iyo2016JACS_1144,PCC1144,Kong2016}. In addition, CaK1144 does not undergo a structural phase transition, sometimes associated with the appearance of internal strain and twinning in these materials. Indeed, one could consider the 1:1 ratio of Ca and K simply as an ordered stoichiometric subsititution of Ca for K at 50 \% ``doping" level. It is thus interesting to compare CaK1144 with hole-doped \bak\ ($x=$0.54 in this study) which has a similar $T_c$ of 34 K \cite{Cho2016BaK} but is randomly disordered on the single (Ba/K) site.

In this paper, the superconducting gap structure of CaKFe$_4$As$_4$ (CaK1144) was studied by measuring the temperature induced variation of the London penetration depth, \dlt, and the  tunneling conductance at low temperatures, both of which probe the density of states (DOS) near the Fermi level, $E_F$. The penetration depth shows saturation at low temperatures and the tunneling spectra exhibit a clear gap in DOS around $E_F$. In-depth data analysis leads to conclusion that CaK1144 has two effective superconducting gaps. The smaller gap is in the range of 1 - 4 meV and the larger gap is between 6-9 meV. The sizeable spread is characteristic of superconductors showing different magnitudes of the superconducting order parameter over the Fermi surface. The larger ratio of the maximum to minimum gap values leads to the overall behavior quite similar to the overdoped \bak\ with $x=0.54$, but different from the optimally - doped \bak\ with $x=0.35-0.4$ where this ratio is about 2 \cite{KimBaK122underdoped,Cho2016BaK}.

\begin{figure}[htb]
\includegraphics[width=8cm]{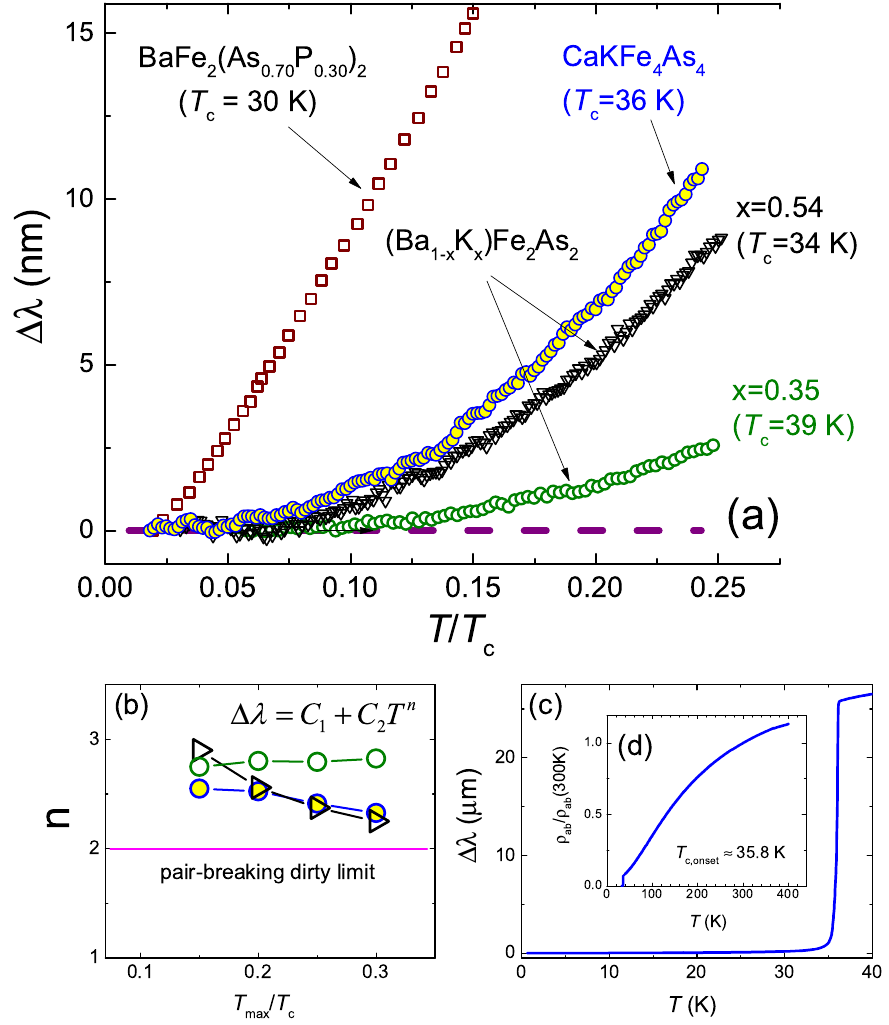}
\centering
\caption{(Color online) (a) $\Delta \lambda$ (T) of CaK1144 (filled circles) compared with other IBS, BaFe$_2$(As$_{0.70}$P$_{0.30}$)$_2$ (nodal gap, $T_c \approx$ 30 K, open squares) \cite{HashimotoCho2012Science_BaP122_dL} and two compositions of \bak\ with $x=$0.35 (no nodes, optimally doped, $T_c \approx$ 39 K, open circles) and $x=$0.54 (no nodes, over-doped, $T_c \approx$ 34 K, open triangles) \cite{Cho2016BaK}. (b) the exponent $n$ obtained from the power law fit, $\Delta \lambda =C_1+C_2 T^n$ as function of the upper fit limit, $T_{max}/T_c$. $n=$2 represents the dirty-limit exponent for the sign-changing order parameters, such as $d-$wave or \spm. Symbols are the same as in (a). (c) Full - temperature range variation of the in-plane London penetration depth $\Delta \lambda (T)$. (d) Normalized in-plane resistivity $\rho_{ab} / \rho_{ab}$ (300 K) showing only superconducting transition.}
\label{fig1}
\end{figure}

Single crystals of CaKFe$_4$As$_4$ were synthesized by high temperature solution growth out of FeAs flux, see Ref.~[\onlinecite{PCC1144}] for details of the synthesis and comprehensive structural, thermodynamic, transport, magneto-optical and spectroscopic characterization. Due to complexity of the growth and potential for unwanted phases, each sample used in the present study was individually screened to be single phase. To this end, the in-plane four-probe resistivity was measured using a \emph{Quantum Design} Physical Property Measurement System (PPMS) in each sample of typical dimensions of approximately 2 $\times$ 0.5 $\times $ 0.02 mm$^3$ and we checked that selected samples showed no extra features except for the superconducting transition, see \fref{fig1}(c). These samples had $R$(300 K)/$R$(40 K) of the order of 15 (compare to 7 of optimally - doped \bak).

The in-plane London penetration depth $\Delta \lambda (T)$ was measured  using a self-oscillating tunnel-diode resonator (TDR) where the sample is subject to a small, 20 mOe, AC magnetic field and the recorded resonant frequency shift from the value of the empty resonator is proportional to sample's magnetic susceptibility, determined by $\lambda$ and sample shape. Detailed description of this technique can be found elsewhere Ref.~[\onlinecite{Prozorov2000PRB,Prozorov2006SST,ProzorovKogan2011RPP}].

For the STM experiment, the sample was mounted onto a sample holder and a piece of brass was glued on top of it. At liquid helium temperature, the sample holder was moved towards a copper beam, lifting off the glued brass piece \cite{Suderow2011,Fente2014} and leaving a freshly cleaved surface for tunneling. The base temperature of this experiment was 800 mK. The energy resolution of the spectroscopy was of 13 $\mu$eV \cite{Suderow2011,Guillamon2008}, well below the temperature induced smearing, which is of order of 70 $\mu$eV. We found flat surfaces, although we did not obtain atomic resolution. A clear signature of the superconducting gap was found consistently over the whole surface, as discussed below. Tunneling conductance, normalized above the superconducting gap was obtained by making a numerical derivative of the tunneling current vs. voltage curves.

\Fref{fig1}(a) shows the low-temperature, $T/T_c \leq$ 0.3, variation of London penetration depth, \dlt, for single crystal CaK1144 compared with three other IBS with comparable $T_c$ values: BaFe$_2$(As$_{0.70}$P$_{0.30}$)$_2$ ($T_c \approx$ 30 K), which exhibits a nodal gap (from our earlier work, Ref.~[\onlinecite{HashimotoCho2012Science_BaP122_dL}]) and two compositions of \bak\ (from our previous work,  Ref.~[\onlinecite{Cho2016BaK}]) with $x=$0.35 exhibiting two isotropic gaps (optimally doped, $T_c \approx$ 39 K) and $x=$0.54 that shows no nodes, but increased anisotropy in at least one of the gaps (over-doped, $T_c \approx$ 34 K). Symbols are described in the caption. \Fref{fig1}(c) shows full - temperature range London penetration depth and normalized resistivity with very sharp transition and no signature of other phases or transitions.

\begin{figure}[htb]
\includegraphics[width=8.2cm]{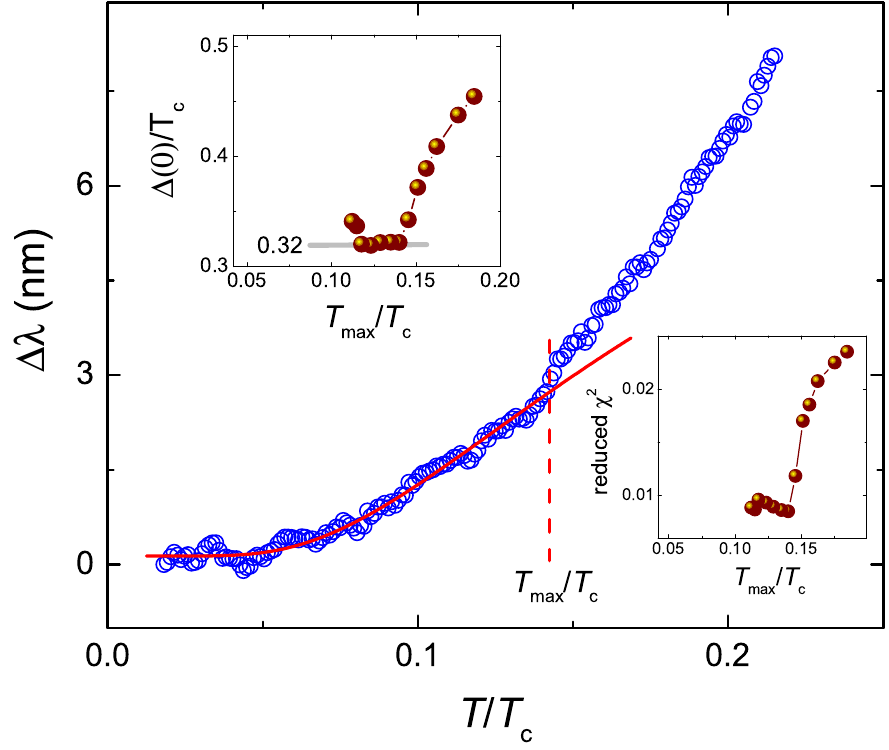}
\centering
\caption{(Color online) A representative BCS fitting with $\Delta/T_c$ as a free fit parameter and fixed $T_{max}/T_c$=0.14. Upper inset: $\Delta/T_c$ obtained from BCS fittings with different $T_{max}/T_c$. Lower inset: reduced $\chi^2$ vs. $T_{max}/T_c$ corresponding to the fitting results shown in the upper inset.}
\label{fig2}
\end{figure}

To numerically characterize the low temperature behavior, we developed a kind of quasiparticle spectroscopy in which we fit the data using different upper limits of the fitting range, $T_{max}$, thus effectively cutting off the quasiparticles with energies exceeding $k_B T_{max}$ (note that throughout the paper we use $k_B=1$). First, in \fref{fig1}(b) we characterize the curvature of \dlt\ by using power - law fitting, $\Delta \lambda = C_1+C_2 (T/T_c)^n$. More details of the procedure are given in our previous study \cite{Cho2016BaK}. The exponent $n=$1 would correspond to the clean limit of line nodes, whereas the exponent $n=$2 is the maximum possible value, reached in the dirty limit of either symmetry - imposed line nodes or sign-changing, but fully gapped, \spm\ pairing. In case of a nodeless $s_{++}$ gap, both clean limit and non-magnetic dirty limit for  either single band or multi - band superconductivity, \dlt\ is  exponential at low temperatures \cite{Hirschfeld1993,Kogan13pairbreaking}. This would correspond to large values of $n >$3-4. We find the values of $n$ clearly exceeding $n=2$ ruling out a nodal gap. Moreover the exponent $n$ vs. $T_{max}/T_c$ in CaK1144 follows almost exactly the behavior found in the overdoped BaK122 ($x=$0.54), which is also seen directly in \fref{fig1}(a). This behavior is consistent with \spm\ pairing with two nodeless gaps \cite{Cho2016BaK}.

To probe the spectroscopic gap in the density of states (which is generally different from the magnitude of the order parameter due to scattering \cite{Kogan2016}) we use the low - temperature Bardeen-Cooper-Schrieffer (BCS) asymptotic behavior expected for the penetration depth,
$\Delta \lambda  = C_1 + C_2\sqrt {\pi \delta/2t} \exp \left(-\delta/t \right)$, where $C_1$, $C_2$ and $\delta  \equiv \Delta \left( 0 \right)/T_c$ and $t=T/T_c$ \cite{ProzorovKogan2011RPP}. \Fref{fig2} shows an example of a good - quality fitting with $T_{max}/T_c=0.14$. By plotting $\delta$ versus the upper fit limit, $T_{max}/T_c$, we expect a saturation when the fit becomes truly exponential indicating a clean gap in the density of states. Indeed, upper inset in \fref{fig2} shows such saturation below $T_{max}/T_c \approx$0.14 at $\delta \approx$ 0.32 $\approx$ 1 meV. Simultaneously, the quality of the fit becomes better and saturates, indicated in the lower inset in \fref{fig2}, by the lowest value of the reduced $\chi^2=\sum(f_{data}-f_{fit})^2/DOF$, where the number of degrees of freedom, $DOF=$(number of data points) - (number of free parameters).

\begin{figure}[htb]
\includegraphics[width=8.5cm]{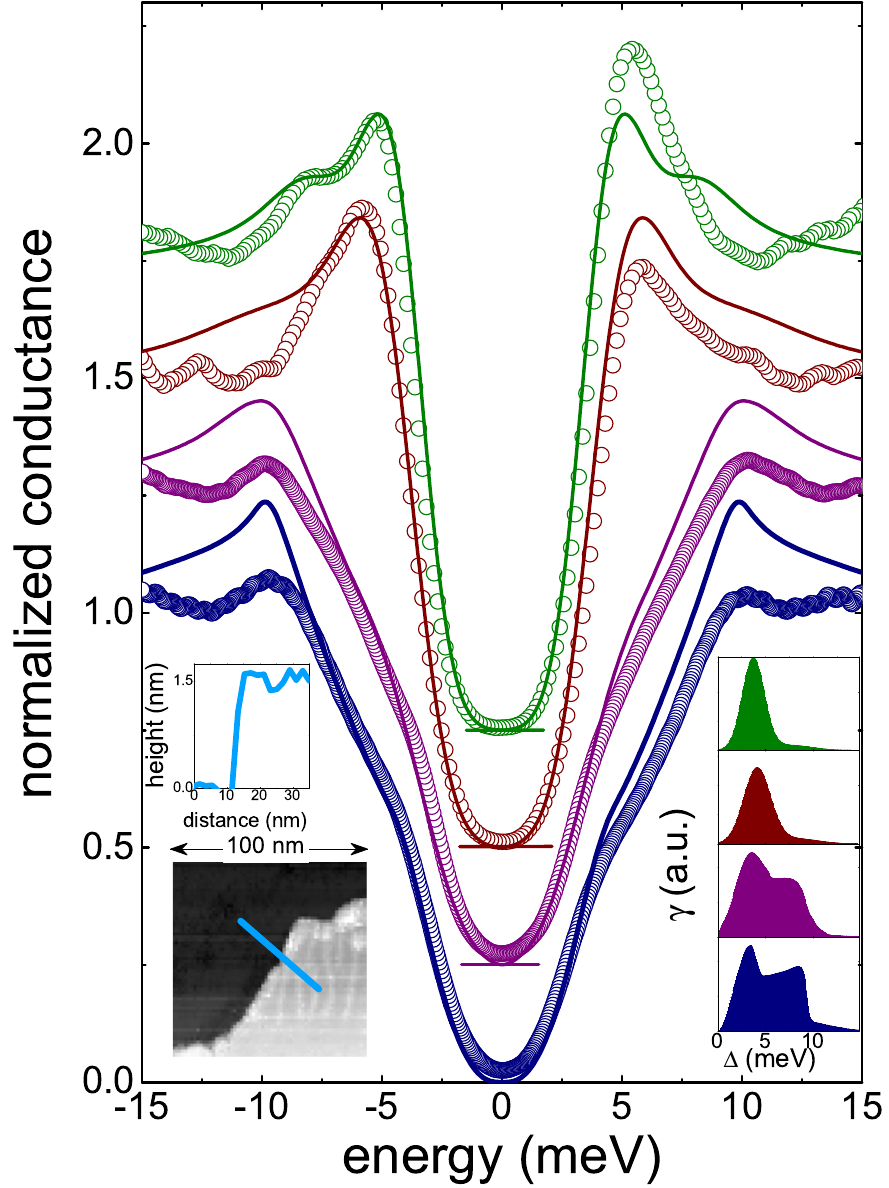}
\centering
\caption{(Color online) Tunneling conductance vs. bias voltage curves measured at 800 mK (symbols) and corresponding fits to BCS theory (solid lines). The curves are shifted, and the zero conductance value is indicated by a line in each curve. These representative results were taken at several points along the line shown in the lower left inset. Left insets show the topography of the surface with a step height of the order of the unit cell height along the c-axis. The size of the image is of 100 $\times$ 100 nm$^2$. The two lower conductance curves are taken on the dark area and the uppermost curves on the bright area. Bottom right insets show the gap distributions used to obtain the fits shown by the lines in the main figure.}
\label{fig3}
\end{figure}

In \Fref{fig3} we show representative tunneling conductance spectra at 800 mK from several locations on the sample surface. All tunneling conductance curves systematically show a negligible density of states close to the Fermi level. Curves, to varying degrees, show two features at about 3 meV (corresponding to 0.54$\Delta(0)/T_c$ if $\Delta(0)/T_c=$1.76, - a single gap weak coupling value) and 8 meV (1.45$\Delta(0)/T_c$). Along a given flat surface, the conductance curves show the same shape, with two clear features. However, when changing the tunneling plane through a roughly unit-cell-high step, the relative size of each feature in the tunneling conductance changes. This shows that the contribution from different parts of the Fermi surface strongly depends on fine details of the surface being tunneled into. To characterize the observed distributions of gap sizes, we have convoluted a density of states of the form $\sum_{\Delta_i}\gamma_i\frac{E}{E^2-\Delta_i^2}$ with the derivative of the Fermi function to obtain the tunneling conductance \cite{Guillamon2008PRL,Rodrigo2004,Suderow2014}. The set of $\Delta_i$, $\gamma_i$ that best reproduces the observed curves gives the lines in \Fref{fig3}. The $\gamma_i$ as a function of the $\Delta_i$ are shown in the lower right inset. The gap distribution shows two peaks around 3 and 8 meV, in agreement with penetration depth results, which, depending on the model, give 2.0-2.4 and 6.0-9.6 meV. There is thus a sizeable spread of gap sizes over the Fermi surface. These values are also in a reasonable agreement with the values inferred from the superfluid density, discussed next.

The superfluid density $\rho_s \equiv (\lambda(0)/\lambda(T))^2$ can be obtained from $\lambda (T)$, provided we can estimate the absolute value of $\lambda(0)$. The TDR technique is suitable for the precision measurements of the changes in the penetration depth \cite{Prozorov2000a}, but not the absolute value. We use two approaches to estimate $\lambda(0)$. First, thermodynamic Rutgers relation is used to estimate Ginzburg-Landau parameter $\kappa_{GL}=\lambda_{GL}/\xi_{GL}$ \cite{Rutgers_Kim2013,PCC1144}:

\begin{eqnarray}
\kappa _{GL}=\sqrt{\frac{T_{c}}{8\pi \Delta C}}\left\vert \frac{\partial H_{c2,c}}{\partial T}\right\vert _{T_{c}}
\label{Rutgers}
 \end{eqnarray}

 \noindent where the jump of the specific heat, $\Delta C=$9.6 J/mol K = 8.32$\times$10$^5$ erg/cm$^3$/K (using molar volume of 115.4 cm$^3$/mol)
the slope of the upper critical field (measured parallel to the $c-$axis) at $T_c$, $dH_{c2,c}/dT=$-4.4$\times$10$^4$ Oe/K  \cite{PCC1144,Kong2016}. Eq.~(\ref{Rutgers}) gives $\kappa _{GL} \approx $60. As shown from the detailed analysis of $H_{c2}(T)$, due to a very short coherence length, $\xi \left( 0\right) =\sqrt{\phi_0/2\pi H_{c2}\left( 0\right)}\approx$ 2.15 nm ($H_{c2}\left( 0\right) \approx $ 71 T), CaK1144 appears to be in the clean limit \cite{PCC1144,Kong2016}. Therefore, we can use clean-limit relation, $\kappa \left( 0\right) =1.206\kappa _{GL}=68.7$ from which $\lambda \left( 0\right) =\xi \left( 0\right) \kappa \left( 0\right) \approx$ 148 nm. Alternatively, we can estimate Ginzburg-Landau $\xi _{GL}=\sqrt{\phi _{0}/2\pi T_{c}\left\vert \frac{\partial H_{c2}}{\partial T}\right\vert _{T_{c}}} \approx$ 1.5 nm, which gives $\lambda _{GL}=\xi _{GL}\kappa _{GL}\approx$ 83 nm. Therefore, $\lambda \left( 0\right) =\sqrt{2}\lambda _{GL}\approx 118 $ nm. These are quite close values resulting a small variation of $\rho_s$ at intermediate temperatures. For the fitting analysis of the superfluid density we use the average of these two value, $\lambda \left( 0\right) =$ 133 nm.

\begin{figure}[htb]
\includegraphics[width=8.5cm]{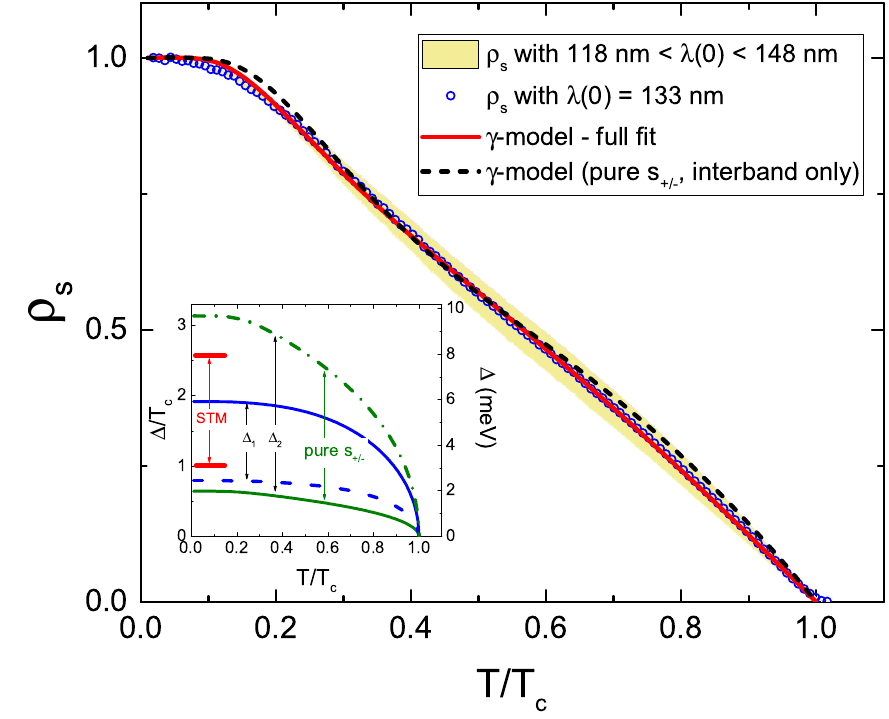}
\centering
\caption{(Color online) Superfluid density, $\rho_s$, calculated with $\lambda (0)=$ 133 nm (symbols). The shaded area shows variation of $\rho_s(T)$ if we used 118 (lower boundary) and 148 nm (upper boundary) found in two different ways to estimate $\lambda (0)$, see text for details. The solid curve is the fit to a self - consistent $\gamma-$model and the dashed curve in interband-only pairing fit.
Inset shows the temperature dependence of the two order parameter values obtained from the fits in the main figure. Lines are for the self-consistent $\gamma-$ model and dash dotted lines for the interband-only fit. The red lines show the energies at which the gap distribution found in STM measurements peaks.}
\label{fig4}
\end{figure}

To further discuss multiband superconductivity, we fit the superfluid density, $\rho_s(T)$, to a two-band $\gamma-$model \cite{Kogan2009gamma}. The two values of the order parameter are calculated self-consistently at each temperature. We use the relative contribution, $\gamma$, from one band (and $1-\gamma$ from the second) as another fit parameter, the total superfluid density is computed. We obtained a very good agreement in the entire temperature range with the order parameters shown in the inset in \fref{fig4}. At very low temperature superfluid density deviates a little, either due to some anisotropy of one of the bands or, more likely, residual scattering. In the fit, we obtained $\Delta _{1}(0)/T_{c} =$ 1.92, $\Delta_{2}(0)/T_{c} =$ 0.64, so that $\Delta_{1}(0)/\Delta_{2}(0) =$ 3.0, which is a factor of 2.0 larger than that found for BaK122 \cite{Cho2016BaK}. In energy units, we obtain $\Delta_{1}(0) =$ 5.92 meV and $\Delta_{2}(0) = $1.97 meV. Furthermore, fitting with interband-only \spm\ model was previously used to analyse the $H_{c2}$ data \cite{Kong2016}. \Fref{fig4} shows our attempt to fit $\rho_s(T)$ to interband - only (pure \spm) model by a dashed line. The result is quite reasonable, although not as good as the full fit described above. Here we obtain two gaps of 2.4 meV and 9.6 meV. As shown in the inset, these values are in a good agreement with 3 meV and 8 meV obtained from the STM experiments.

In the case of $\gamma-$ model fitting, the main uncertainty comes from insufficient information on the real electronic band structure. Calculations show three hole-like bands and two electron-like bands with distinct 3D character of the outer hole-like sheet  \cite{ARPES2016}. Partial densities of states, Fermi velocities (or better plasma frequencies) are required as input. For the present analysis we fixed equal DOS on each effective band, $n_1=n_2=$ 0.5 and obtained from the fitting, $\lambda_{22}=$0.35, $\lambda_{12}=$1.13 and $\gamma=$ 0.56. For the full fit we chose $\lambda_{11}=$0.85 to produce correct transition temperature assuming characteristic energy scale of a superconducting interaction of the order of Debye temperature $T_D=$300 K (determined as a fitting parameter from specific heat measurements \cite{PCC1144}). Here, $\lambda_{ij}$ is the symmetric interaction matrix (we show absolute values here), see Ref.~\onlinecite{Kogan2009gamma,ProzorovKogan2011RPP} for details. For pure \spm, interband-only pairing, we obtained $\lambda_{11}=\lambda_{22}=$0, $\lambda_{12}= $1.55, $\gamma=$ 0.65 and $n_1=$ 0.89. These two fits give limiting cases and will be refined when electronic bandstructure parameters are available. Importantly, we demonstrate that our data are described by two effective superconducting gaps very well. The fitting shows significant interband pairing strength and quantitatively agrees with independent STM studies. Recent angle-resolved photoemission spectroscopy finds larger than BCS weak-coupling limit superconducting gaps, isotropic in the $ab-$plane \cite{ARPES2016}. More detailed studies in the rest of the Brillouin zone for all Fermi surface sheets are needed to fully determine all the gaps, especially the small ones, which must exist to reconcile with our observations.

In conclusion, precision measurements of the London penetration depth and low temperature STM spectroscopy show unambiguously a fully gapped multiband superconductivity in single crystals of CaKFe$_4$As$_4$. Analysis with two effective gaps gives small gap in the range of 1-4 meV and the large gap is between 6-9 meV. The overall behavior is quite similar to optimally doped \bak\ with $x=0.54$.
Notably, while the overall spread of gap values (mostly given by the difference between the averages of the two ranges of gap sizes) is lower in the presence of substitutional disorder (i.e., in BaK122), the \spm\ physics with two effective gaps is clearly present in stoichiometric and substituted systems with high $T_c$'s.

We thank A. Gurevich, D. D. Johnson, A. Kaminski, V.~G.~Kogan and Lin-Lin Wang for useful discussions. This work was supported by the U.S. Department of Energy (DOE), Office of Science, Basic Energy Sciences, Materials Science and Engineering Division. Ames Laboratory is operated for the U.S. DOE by Iowa State University under contract DE-AC02-07CH11358. The work in Madrid was supported by the Spanish Ministry of Economy and Competitiveness (FIS2014-54498-R and MDM-2014-0377), by the Comunidad de Madrid through program Nanofrontmag-CM (S2013/MIT-2850) by Axa Research Fund, FP7-PEOPLE-2013-CIG 618321 and the European Research Council (grant agreement 679080). Madrid's group also acknowledges SEGAINVEX workshop of UAM.

\bibliographystyle{apsrev4-1}

%

\end{document}